\documentclass[12pt]{article}

\usepackage{amssymb,amsmath,amsbsy,amsthm}
\usepackage{graphicx}
\usepackage{authblk}

\newcommand{\eB}{e_{0}}
\newcommand{\eR}{e_{R}}

\newcommand{\pdag}{^{\phantom\dag}}
\newcommand{\ud}{\mathrm{d}}
\newcommand{\ta}{\tilde{a}}
\newcommand{\xxa}{\stackrel {\scriptscriptstyle \times}{\scriptscriptstyle \times} \!}
\newcommand{\xxe}{\! \stackrel {\scriptscriptstyle \times}{\scriptscriptstyle \times}}
 
\newcommand{\nna}{\stackrel {\scriptscriptstyle \circ}{\scriptscriptstyle \circ}\! }
\newcommand{\nne}{\! \stackrel {\scriptscriptstyle \circ}{\scriptscriptstyle \circ}}

\newcommand{\ee}{\,{\rm e}}
\newcommand{\ii}{{\rm i}}
\newcommand{\ve}{{\bf e}}
\newcommand{\he}{\hat e}

\newcommand{\vp}{{\bf p}}
\newcommand{\vx}{{\bf x}}
\newcommand{\vy}{{\bf y}}

\newcommand{\Ref}[1]{(\ref{#1})}

\title{\Large{\bf{Gauge invariance, correlated fermions, and\\photon mass in 2+1 dimensions}}}

\date{\vspace{-1.0cm}\small 
October 30, 2013 
\vspace{0.2cm}}

\renewcommand\Authfont{\scshape\normalsize}
\renewcommand\Affilfont{\itshape\small}
\author[1,*]{Jonas de Woul}
\author[1,\dag]{Edwin Langmann}
\affil[1]{Department of Theoretical Physics, Royal Institute of Technology KTH\newline SE-106 91 Stockholm, Sweden \vspace{2mm}}

\begin{document}

\maketitle

\let\oldthefootnote\thefootnote
\renewcommand{\thefootnote}{\fnsymbol{footnote}}
\footnotetext[1]{Electronic address: {\tt jodw02@kth.se}}
\footnotetext[2]{Electronic address: {\tt langmann@kth.se} (Corresponding author; tel: +46-8-55378173)}
\let\thefootnote\oldthefootnote

\vspace{-1.5cm}

\begin{abstract}
We present a 2+1 dimensional quantum gauge theory with correlated fermions that is exactly solvable by bosonization. This model describes a system of Luttinger liquids propagating on two sets of equidistant lines forming a grid embedded in two dimensional continuum space; this system has two dimensional character due to density-density interactions and due to a coupling to dynamical photons propagating in the continuous embedding space. We argue that this model gives an effective description of partially gapped fermions on a square lattice that have density-density interactions and are coupled to photons. Our results include the following: after non-trivial renormalizations of the coupling parameters, the model remains well-defined in the quantum field theory limit as the grid of lines becomes a continuum; the photons in this model are massive due to gauge-invariant normal-ordering, similarly as in the Schwinger model; the exact excitation spectrum of the model has two gapped and one gapless mode. 

 \end{abstract}

\section{Introduction}
There are several example in the history of physics where an exactly solvable model, which first appeared in one area of physics, was later found useful in another. For example, recent developments in optical lattices have led to experimental realizations of one dimensional exactly solvable quantum many body models that, for many years, were regarded as mere toy models by many physicists (see e.g.\  \cite{ColdAtoms}). Another example is the massless Thirring model \cite{Thirring}, which was originally proposed as a toy model in elementary particle physics but later, through the work of Luttinger \cite{Luttinger} and Haldane \cite{Haldane} among others, found an important application in condensed matter physics as a realistic prototype model for one dimensional correlated fermions (an earlier pioneering paper on this is by Tomonaga \cite{Tomonaga}). We thus hope that the results presented in this paper are of some general interest beyond the motivations which we provide. 

We present a quantum field theory model describing interacting fermions coupled to a dynamical $\mathrm{U}(1)$-gauge field in two dimensions (2D) which is exactly solvable. We also obtain exact results for this model. As will be discussed,  our motivation are 2D correlated fermions systems like the cuprate superconductors \cite{BednorzMueller}, but our model is probably too simple to provide a realistic description of a material existing in nature. However, we believe it still is valuable: First, it is close enough to realistic such models to serve as test case for pertinent approximation methods; second, it provides a simple pedagogical example for multiplicative renormalization in quantum field theory; third, it provides a 2+1 dimensional example for dynamical photon mass generation due to an anomaly, similarly as in the {\em Schwinger model} \cite{Schw3}  (this is further discussed in our conclusions); fourth, it is an example of a quantum field theory model which can be defined and studied in a way which is simple and mathematically precise at the same time.
 
Our starting point is a model first proposed by Mattis to describe 2D fermions with a square Fermi surface and density-density interactions, and which can be solved exactly by bosonization \cite{Mattis};  see also \cite{Hlubina,Luther,FSL,SL,VM} for interesting work on similar models in the condensed matter literature. In  recent work we presented physics arguments that the Mattis model provides an effective description of fermions on a square lattice with local interactions in a partially gapped phase away from half filling \cite{EL1,EL2,dWL1}. We also showed that it is possible to define and solve the Mattis model rigorously according to the standards of mathematical physics \cite{dWL2}. 

The model studied in this paper is obtained by coupling the Mattis model to a dynamical $\mathrm{U}(1)$-gauge field using the {\em gauge principle}, i.e.\ coupling the fermions to a dynamical photon field so that local gauge invariance is preserved. We find that the gauge principle has an interesting twist: normal ordering, which is needed to make the Mattis model well-defined, breaks gauge invariance. For this reason, minimal coupling alone does not give a gauge invariant result; one also has to use gauge invariant normal ordering \cite{Schw1}, and this leads to additional fermion-photon coupling terms (note that the latter arise not only from the kinetic- but also from the interaction terms of the fermion Hamiltonian).  

We believe that it should be possible to also derive this model as an effective description of $\mathrm{U}(1)$-lattice gauge theory coupled to interacting lattice fermions, in generalization of our results in \cite{EL2,dWL1}. However, such derivations are notoriously difficult, and it is common to regard the gauge principle as a reliable substitute for such a derivation. As a well-known example we recall the Standard Model in elementary particle physics which, in principle, should be derivable from a more fundamental theory (e.g.\ String theory). However, this is very difficult in practice \cite{Raby}; instead, one uses the gauge principle which, for a given model of matter and gauge group, fixes the gauge theory completely. We admit that the gauge principle does not provide a proof: we cannot rule out the possibility that the true effective model for this lattice gauge theory is more complicated, with additional terms that spoil the exact solvability; the exact solution of the model presented in this paper should be a useful starting point even in that case.

As mentioned, the solution method we use is bosonization. There is a large literature on bosonization in higher dimensions motivated by 2D correlated fermions systems;  see e.g.\ Ref.\ \cite{HKM} for review. The idea of mapping fermions to free bosons has been used before to study 2D interacting fermions coupled to gauge fields \cite{KHR,KHM}; see e.g.\ \cite{MCD}, Sec.~9 for review (other work on the possible role of $\mathrm{U}(1)$-gauge fields for 2D correlated fermion systems is discussed in Section~\ref{sec5}). However, our work differs in important details from all previous work we are aware of. In particular, our results are exact. Moreover, we clearly identify a model which can be defined and studied rigorously according to the standards of mathematical physics. This is due to the following: it is possible to formulate and bosonize the model exactly with particular UV cutoffs in place, and it therefore is possible to check by simple means that all definitions and the UV limit are well-defined mathematically. However, to simplify our presentation, we omit in this paper some technical details which would be of interest mainly to readers interested in mathematical rigor (we plan to provide these details elsewhere). In particular, the equivalence of the fermion- and boson gauge theory models presented in this paper is an exact result. 

Depending on background and interest, the paper can be read in different ways: We state our definitions and results using a formal notation which, as we hope, should allow readers not interested in technical details to quickly asses our results; readers interested in technical details can consult Ref.\ \cite{dWL2} for a dictionary to translate our formal notation into mathematically precise statements (the latter is only described in words in the present paper). More pragmatic readers can regard our derivation of the bosonized model given in the beginning of Section~\ref{sec4} as heuristics, and to take this boson model as starting point: its key properties can be easily checked without knowing how it is derived (as will be explained), and since it is  a {\em free} quantum field theory (i.e.\ there are no non-linear interactions), it is easy to make it mathematically precise (mathematical details on how to make a free boson quantum field theory mathematically precise are collected in \cite{dWL2}, Appendix~B.1, for example). Still other readers might be only interested in a gauge theory coupled to bosonic matter which is free and with massive photons; such readers can start with Eq.\ \Ref{cLGM} where the Lagrangian defining this model is given (it is easy to generalize this example to other dimensions or construct a variant which is rotation invariant). 

Our plan is to give a concise presentation and discussion of our results in the main text (Section~\ref{sec2}--\ref{sec5}), and to defer technical details to six appendices.  Section~\ref{sec2} discusses the Mattis model and its proposed relation to lattice fermions (the latter part can be skipped without loss of continuity). In Section~\ref{sec3} we present our gauge theory model and some mathematical results that we need. The exact solution of  this model is given in Section~\ref{sec4}. Our conclusions are in Section~\ref{sec5}. In Appendix~\ref{App_formal} we derive the Hamiltonian formulation from the Lagrangian formulation of our model. Appendix~\ref{App_normalorder} provides details on gauge invariant normal ordering. Formulas that provide an independent check of gauge invariance are spelled out in Appendix~\ref{App_gauge}. Computational details on the diagonalization of the model Hamiltonian, the functional integral formalism that we use to find response functions, and our proof of the Meissner effect can be found in Appendices~\ref{App_solution}, \ref{App_Lagrange} and \ref{App_Meissner}, respectively. 

\noindent {\bf Notation}: $\mu,\nu=0,+,-$ are space-time indices, $s,s'=+,-$  space indices, $r,r'=\pm$ chirality indices;  the space-time metric signature is $(-,+,+)$; $x=(x^0,\vx)$, $x^0=ct$, are space-time 
coordinates with 2D positions $\vx =(x_+,x_-)$ and time $t$; $c$ is the velocity of light; $\partial_s={\partial}/{\partial x_s}$ are spatial derivatives; $\psi^{(\dag)}_{r,s}$ are standard fermion field operators; $A_\mu$ is the gauge potential, $E_s$ are the electric field components, and $B=\partial_+A_- - \partial_- A_+$ is the magnetic field. Common argument $\vx$ of field operators are suppressed whenever possible without danger of confusion.

\section{Fermion model}
\label{sec2}
The Hamiltonian defining the Mattis model can be written in position space as follows,  
 \begin{equation}
\label{Mattis Hamiltonian}
\begin{split} 
H_M = \sum_{r,s=\pm} \int \ud^2 x\, \Bigl(  rv_F :\! \psi^\dag_{r,s}(-\ii \partial_s)\psi_{r,s}\pdag\!: + \sum_{r',s'=\pm} g_{r,s,r',s'} :\!\psi^\dag_{r,s}\psi_{r,s}\pdag\!: :\!\psi^\dag_{r',s'}\psi_{r',s'}\pdag\!:  \Bigr)
\end{split} 
\end{equation}
with $\psi_{r,s}^{(\dag)}(\vx)$ fermion field operators satisfying the usual canonical anticommutator relations $\bigl\{ \psi_{r,s}^{\phantom\dag}(\vx),\psi_{r',s'}^\dag(\vy)\bigr\}=\delta_{r,r'}\delta_{s,s'}\delta^2(\vx-\vy)$, etc., and colons denoting  standard fermion normal ordering with respect to a Dirac vacuum \cite{dWL2}. These definitions are formal since  important UV regularizations are suppressed:\ while the $x_s$-component of $\vx$ in $\psi^{(\dag)}_{r,s}(\vx)$ is continuous, the $x_{-s}$-component is discretized to integer multiples of a UV cutoff $\tilde a$, and thus the integrals and Dirac deltas have to be interpreted as partial Riemann sums and Kronecker deltas, respectively \cite{dWL2}. Moreover, the interaction potential, which in \Ref{Mattis Hamiltonian} corresponds to $g_{r,s,r',s'}\delta^2(\vx-\vy)$, is regularized by replacing the Dirac delta by a suitable smeared delta function.\footnote{To be more specific: $\delta^2(\vx)$ is replaced by $(1/\ta^2) f(\vx/\ta)$ with a particular smooth function $f$  such that $\int\ud^2 x\,  f(\vx)=1$ \cite{dWL2}.} The coupling constants scale with the UV cutoff as  
\begin{equation}
g_{r,s,r',s'} = \ta\pi v_F\bigl(\gamma_1\delta_{s,s'}\delta_{r,-r'}+ \gamma_2\delta_{s,-s'}/2\bigr), 
\end{equation} 
with the Fermi velocity $v_F>0$ and dimension-less constants $\gamma_{1,2}$ such that $|\gamma_1|<1$ and $|\gamma_2|<|1+\gamma_1|$. The above scaling of the coupling constants is not only obtained by deriving the model from lattice fermions \cite{EL2}, but it also ensures that $H_M$ has a non-trivial limit as $\ta\to 0$ \cite{dWL2}. The restrictions on $\gamma_{1,2}$ are to ensure stability of the Dirac vacuum \cite{dWL2}. 

The Hamiltonian in \Ref{Mattis Hamiltonian} describes, at face value, a system of Luttinger liquid living on two sets of parallel lines forming a 2D grid as follows,\footnote{Note that the following picture is rotated by an angle $\pi/4$ as compared to the one in Figure~\ref{FermiSurface} below.}
\begin{center} 
\setlength{\unitlength}{2mm}
\begin{picture}(20, 20)
  \linethickness{0.15mm}
  \multiput(0, 0)(1, 0){21}{\line(0, 1){20}}
  \multiput(0, 0)(0, 1){21}{\line(1, 0){20}}
\end{picture}
\end{center} 
with the UV cutoff $\ta$ equal to the distance of two adjacent parallel lines (as indicated in the figures, we also use a IR cutoff $L$ corresponding to the extension of space in the $x_+$- and $x_-$-directions \cite{dWL2}). The fermions $\psi_{r,s}(x_+,x_-)$ propagate on the horizontal and vertical lines for $s=+$ and $-$, respectively, i.e., $x_s$ is continuous and $x_{-s}$ a discrete label. This system of one dimensional Luttinger liquids has a two dimensional character due to density-density interactions between fermions on different lines. In the rest of this section we shortly describe another physical interpretation of the Mattis model as an effective description of 2D lattice fermions in a partially gapped phase (one can ignore this without loss of continuity).

We describe the relation of the Mattis model to lattice fermions using Figure~\ref{FermiSurface},  which shows the Brillouin zone corresponding to a square lattice (dashed large square) divided into regions of different sizes.\footnote{Note that this figure is rotated by an angle $\pi/4$ as compared to the grid drawn above; e.g.\ �the arrow is parallel with the $k_+$-direction.}  Close to half filling, mean field theory indicates that the system is partially gapped, and there is an underlying Fermi surface in the so-called {\em nodal} regions (the four tilted rectangles) \cite{dWL1}. We model this underlying surface by straight arcs, which either corresponds to a truncated Fermi surface or the portion of a closed Fermi pocket having dominant momentum occupation. The Mattis model describes the fermion degrees of freedom in the vicinity of these arcs. It is written in terms of four fermion field operators $\psi_{r,s}$, $r,s=\pm$, in one-to-one correspondence with the nodal regions. The quantum field theory limit making this model amenable to bosonization involves removing the momentum cutoff orthogonal to the arcs (indicated by the arrow in the nodal $(+,+)$-region, for example), which is possible after normal ordering \cite{EL2}. The arc-picture underlying our derivation of the Mattis model is supported by renormalization group studies of weakly coupled 2D Hubbard-like systems \cite{FRS}. It is also a signature feature of the pseudogap phase as observed in angle-resolved photoemission experiments on hole-doped cuprates \cite{ARPES}. Further details, including  computation results in support of this interpretation, can be found in Refs.\ \cite{EL2,dWL1}.

\begin{figure}[t]
\centering
\includegraphics[width=0.7\textwidth, clip=true, trim= 8cm 1.0cm 6cm 1.5cm]{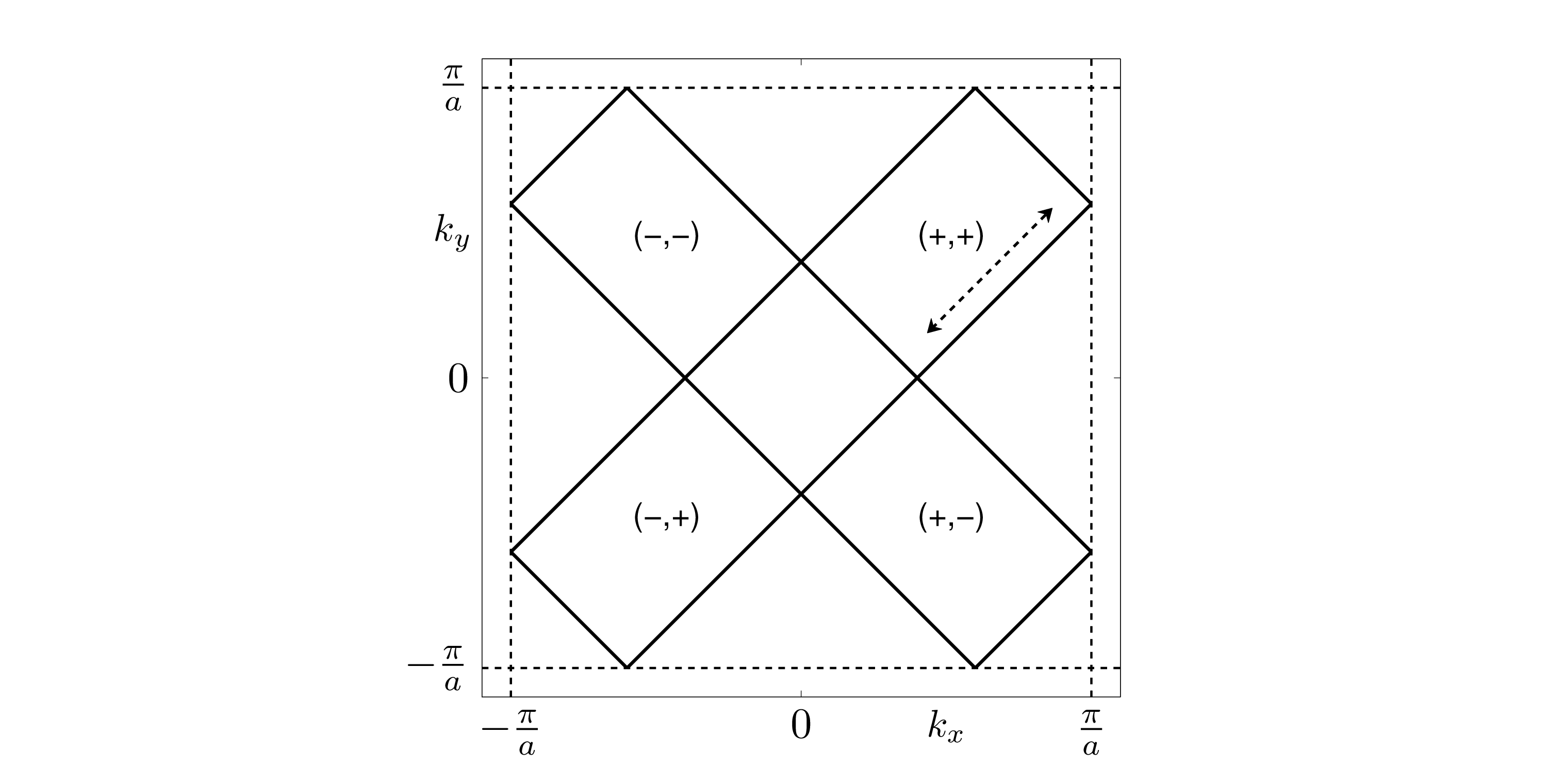}
\caption{\label{FermiSurface} 
Brillouin zone of a square lattice. Our model provides a low-energy description of the fermions with momenta in one of the four tilted rectangles labeled by $(r,s)$, $r,s=\pm$. Fermion degrees of freedom outside the rectangles are assumed to be gapped.} 
\end{figure}

\section{Definition of the model}
\label{sec3} 
The Hamiltonian of our quantum gauge theory model is obtained by coupling the Mattis Hamiltonian in \eqref{Mattis Hamiltonian} to a dynamical electromagnetic field,
\begin{equation}
\label{Gauged Mattis Hamiltonian}
\begin{split}
H = \sum_{r,s=\pm} \int \ud^2 x\, \Bigl(  rv_F \nna \psi^\dag_{r,s}(-\ii \partial_s + \eB A_s)\psi_{r,s}\pdag\nne 
+ \sum_{r',s'=\pm} g_{r,s,r',s'} \nna\psi^\dag_{r,s}\psi_{r,s}\pdag\nne\nna\psi^\dag_{r',s'}\psi_{r',s'}\pdag\nne  \Bigr)
\\
+ \frac12\int \ud^2x \xxa\Bigl(E_+^2 + E_-^2 + c^2 B^2\Bigr)\xxe
\end{split}
\end{equation}
with $[A_s(\vx),E_{s'}(\vy)]=\ii\delta_{s,s'}\delta^2(\vx-\vy)$, etc., $e_0$ the bare charge,  $\xxa\cdots\xxe$ boson normal ordering, and $\nna\cdots\nne$ a gauge-invariant generalization of fermion normal ordering (see below). The Gauss law operators generating gauge transformations $A_s\to A_s + \partial_s\chi$, etc., are
\begin{equation}
\label{Gauss law generator}
G[\chi]=\int \ud^2x\, \chi \sum_{s=\pm}\Bigl(-\partial_sE_s + \eB\sum_{r=\pm}\nna\psi^\dag_{r,s}\psi_{r,s}\pdag\nne\Bigr),
\end{equation} 
and the physical states are those annihilated by $G[\chi]$, for arbitrary real-valued  and differentiable functions $\chi(\vx)$. Note that, except for the normal-ordering procedures, the Hamiltonian in \eqref{Gauged Mattis Hamiltonian} is obtained from the Mattis Hamiltonian by standard minimal coupling;  see Appendix~\ref{App_formal} for details.  We note in passing that the gauge field operators are well-defined without UV regularization in addition to normal ordering (details will be spelled out elsewhere). 

As already indicated, fermion normal ordering plays a key role for our model. Indeed, formulating a sensible quantization of the classical theory is non-trivial, even in the absence of gauge fields: In order for the Hamiltonian to be bounded from below (have a ground state), we need to normal-order all fermion bilinears with respect to the Dirac vacuum in which all negative energy states are occupied. An important consequence of normal-ordering is that the fermion densities $J_{r,s} \equiv \, :\!\psi^\dag_{r,s}\psi^{\phantom\dag}_{r,s}\!:$ obey the anomalous commutator relations \cite{Schw1} 
\begin{equation}
\label{JJ}
[J_{r,s}(\vx),J_{r',s'}(\vy)]=r\delta_{r,r'}\delta_{s,s'}({2\pi\ii\tilde a})^{-1} \partial_s\delta^2(\vx-\vy),
\end{equation}
and the non-interacting part of \eqref{Mattis Hamiltonian} can be expressed in terms of these densities using the operator identity
\begin{equation}
\label{Kronig identity}
\int \ud^2x\, :\!\psi^\dag_{r,s}r \left(- \ii  \partial_s\right)\psi^{\phantom\dag}_{r,s}\!: \, = \pi\ta\int \ud^2 x  \xxa J_{r,s}^2 \xxe
\end{equation}
(see \cite{dWL2}, Proposition 2.1, for proofs of these statements). The commutation relations in \eqref{JJ} imply that
\begin{equation}
\label{bosons} 
\begin{split}
\partial_s\Phi_s &= \sqrt{\pi\tilde a}\bigl( J_{+,s}+ J_{-,s} \bigr)
\\
\Pi_s &= \sqrt{\pi\tilde a}\bigl(-J_{+,s} + J_{-,s}\bigr)
\end{split}
\end{equation}
define boson operators obeying the usual canonical commutator relations $[\Phi_s(\vx),\Pi_{s'}(\vy)]=\ii\delta_{s,s'}\delta^2(\vx-\vy)$, etc. It follows, using \eqref{Kronig identity}, that the Mattis Hamiltonian in \eqref{Mattis Hamiltonian} can be written in terms of free bosons, and this is the key step towards the exact solution of the Mattis model \cite{dWL2}.

The gauge fields are quantized as usual by a partial gauge fixing $A_0=0$, postulating the canonical commutation relations (see below \eqref{Gauged Mattis Hamiltonian}), and imposing the Gauss law operator constraint on the Hilbert space \cite{Jackiw}. However, the corresponding quantum Hamiltonian is {\em not} obtained as the straightforward quantization of the minimally coupled classical Hamiltonian: due to the anomalous commutators in \eqref{JJ}, the Gauss law operators $G[\chi]$ would in this case no longer commute with the Hamiltonian, and the theory would thus not be gauge invariant. The remedy of this problem is to introduce a manifestly gauge-invariant normal-ordering prescription. We use the point-splitting method pioneered by Schwinger \cite{Schw1}: start with the gauge-invariant expression
\begin{equation} 
\label{BL}
\psi^\dag_{r,s}(\vx-\epsilon\ve_s/2) \exp\left( \ii \eB\int_{-\epsilon/2}^{\epsilon/2} A_s(\vx+\xi\ve_s)\ud\xi\right) \psi^{\phantom\dag}_{r,s}(\vx+\epsilon\ve_s/2),
\end{equation} 
including a line integral of the gauge field, and define gauge-invariant fermion normal ordering of bilinears, $\nna\psi^\dag_{r,s}\psi^{\phantom\dag}_{r,s}\nne$, as the limit $\epsilon\to 0$ of \eqref{BL} after first subtracting off its singular part. The result is
\begin{equation}
\label{JrsA} 
\nna \psi^\dag_{r,s}\psi^{\phantom\dag}_{r,s}\nne \; =  J_{r,s} + r \eB A_{s}/({2\pi\tilde{a}}),
\end{equation}
and similarly
\begin{equation}
\label{gauge-invariant fermion normal-ordering with derivative}
\begin{split}
\int \ud^2x\, \nna \psi_{r,s}^\dag r\left( { - \ii \partial _s } + \eB A_s\right)\psi_{r,s}\pdag \nne = \int \ud^2 x\, \Bigl( :\! \psi_{r,s}^\dag r ( { - \ii \partial _s })\psi_{r,s}\pdag \!: +r\eB A_s J_{r,s}  + \frac{\eB^2}{4\pi\ta} A_s^2 \Bigr)
\end{split}
\end{equation}  
(computational details can be found in Appendix~\ref{App_normalorder}). An important feature of \eqref{gauge-invariant fermion normal-ordering with derivative} is the "bare" photon mass term in the second line; as noted, this is a direct consequence of gauge-invariant normal-ordering.  

We stress that our derivations of \Ref{JrsA}  and \Ref{gauge-invariant fermion normal-ordering with derivative} are not mathematically rigorous but rather physical motivations which can be ignored without loss of continuity: the reader can regard  \Ref{JrsA}  and \Ref{gauge-invariant fermion normal-ordering with derivative}  as definitions.  It is easy to check that they  are appropriate: the expressions on the r.h.s. of these equations are mathematically well-defined since standard normal ordering is, and that they define gauge invariant quantities can be checked by simple computations; see Appendix~\ref{App_gauge} for details. 

Inserting  \Ref{JrsA}  and \Ref{gauge-invariant fermion normal-ordering with derivative}  in \Ref{Gauged Mattis Hamiltonian} one obtains a formula expressing the model Hamiltonian in terms of fermion operators defined with standard normal ordering. One thus can use \Ref{Kronig identity} and \Ref{bosons} to bosonize this Hamiltonian.

\section{Solution} 
\label{sec4} 
The Hamiltonian and the Gauss law operators of the gauged model are now bosonized using the above results:
\begin{equation}
\begin{split}
\label{HGM}
H = \frac12\int \ud^2x \xxa\Bigl(v_F\sum_{s=\pm}\bigl[(1-\gamma_1)(\Pi_s-\eR A_s)^2+(1+\gamma_1)(\partial_s\Phi_s)^2
\\
  +\gamma_2(\partial_s\Phi_s)(\partial_{-s}\Phi_{-s})\bigr] + E_+^2 + E_-^2 + c^2 B^2\Bigr)\xxe
\end{split}
\end{equation} 
and 
\begin{equation} 
\label{bG} 
G[\chi] = \int \ud^2x\, \chi \sum_{s=\pm} \partial_s(-E_s+\eR\Phi_s), 
\end{equation} 
with the {\em renormalized} charge 
\begin{equation}
\eR = \frac{\eB}{\sqrt{\pi \tilde a}}. 
\end{equation} 
Note that the scaling of the model parameters and boson fields are such that our gauge theory model remains well-defined in the UV limit $\ta \to 0^+$. The charge renormalization $\eB\to \eR$ shows that photons can have stronger influence on physical properties than what superficial arguments might suggest.  Under a gauge transformation, $A_s\to A_s+\partial_s\chi$, $E_s\to E_s$, $\Phi_s\to \Phi_s$ and $\Pi_s\to\Pi_s+\eR \partial_s\chi$, such that $\Pi_s-\eR A_s$, and thus $H$ in \eqref{HGM}, are manifestly gauge invariant. 

It is remarkable that, after bosonization and charge renormalization, the UV cutoff can be easily removed; since we have suppressed details of the UV regularization, this is not explicitly visible here. We therefore note that, in the following, the removal of the UV cutoff is understood. 

The Hamiltonian in \eqref{HGM} is quadratic in boson operators and can therefore be diagonalized by a Bogoliubov transformation  (in the following we outline this computation and state the main results; further details are given in Appendix~\ref{App_solution} and \cite{dWL2}, Appendix~C).  To this end, we perform a Fourier transformation, $E_s(\vx)\to \hat E_s(\vp)$, and define longitudinal- and tranverse fields by $|\vp|\hat E_L(\vp) = \ii p_+\hat E_+(\vp)+\ii p_-\hat E_-(\vp)$ and $|\vp|\hat E_T(\vp) = \ii p_+\hat E_-(\vp)-\ii p_-\hat E_+(\vp)$ (similarly for $\hat A_s$, $\hat \Pi_s$, and $\hat \Phi_s$). The Gauss law constraint then implies that $|\vp|(-\hat E_L+\eR\hat\Phi_L)$ is zero on the physical space. Fixing the Coulomb gauge, $\hat A_L=0$, and solving the Gauss law, $\hat E_L=\eR\hat\Phi_L$, we obtain the gauge-fixed Hamiltonian $H_{g.f.}$ describing transverse photons ($\hat A_T$) coupled to longitudinal- and transverse  plasmons ($\hat\Phi_L$ and $\hat\Phi_T$, respectively). This Hamiltonian can be written in the diagonal form 
 \begin{equation} 
 H_{g.f.}=E_0+\sum_{j=1,2,3}\sum_{\vp} \omega_j(\vp)b^\dag_j(\vp)b\pdag_j(\vp)
 \end{equation} 
 with  creation- and annihilation operators $b^\dag_j(\vp)$ and $b\pdag_j(\vp)$, respectively, obeying the usual relations $[b_j(\vp),b^\dag_{j'}(\vp')]=\delta_{j,j'}\delta_{\vp,\vp'}$ etc., and the groundstate energy $E_0$. The exact dispersion relations $\omega_j(\vp)$ are computed from the eigenvalues of a certain $3\times 3$ matrix   (this matrix is constructed in Appendix~\ref{App_solution}).   We obtain the following characteristic polynomial of this matrix whose zeros $\lambda=\lambda_j$ are equal to $\omega_j(\vp)^2$,
\begin{equation}
\label{cp}  
\begin{split} 
\lambda(\lambda-\Theta^2-c^2|\vp|^2)(\lambda-\Theta^2 -\tilde v_F^2 |\vp|^2) 
 +|\vp|^4S^2 v_-^2 ( v_+^2(\lambda-c^2|\vp|^2)-c^2\Theta^2 )
\end{split} 
\end{equation} 
with $v_\pm^2 = v_F^2(1-\gamma_1)(1+\gamma_1\pm\gamma_2)/2$, $\tilde v_F^2 = v_F^2(1-\gamma_1^2)$, $\Theta=\sqrt{v_F(1-\gamma_1)}|\eR|$, and $S=|\sin(2\varphi)|=|2p_+p_-|/|\vp|^2$  ($|\vp|^2\equiv p_+^2+p_-^2$). This allows us to compute the exact dispersion relations of the model (see Appendix~\ref{App_solution} for details). We obtain two gapped modes with the same gap proportional to the renormalized charge: $\omega_1(\boldsymbol{0})=\omega_2(\boldsymbol{0})=\Theta$, whereas the third mode is gapless: $\omega_3(\boldsymbol{0})=0$. Moreover, for $c\gg v_F$ and $v_F^2|\vp|^2S\ll 1$,   $\omega_1(\vp)\approx \sqrt{\Theta^2+c^2|\vp|^2}$, $\omega_2(\vp)\approx \sqrt{\Theta^2+\tilde v_F^2|\vp|^2}$, and 
\begin{equation}
\label{omega3}
\omega_3(\vp) \approx |\vp|^2S\sqrt{\frac{c^2v_-^2(\Theta^2+ v_+^2|\vp|^2)}{(\Theta^2+c^2|\vp|^2)(\Theta^2 + \tilde v_F^2 |\vp|^2)}}.
\end{equation}
(more precise formulas are given in \Ref{om1om2} and \Ref{om3}).  This shows that $\omega_1$ gives the energy spectrum of the dressed transverse photon, whereas $\omega_2$ and $\omega_3$ give those of the dressed transverse- and longitudinal plasmons, respectively. Remarkably, the behavior of the  last mode is {\em qualitatively} different for $\eR=0$ and $\eR\neq 0$: in the former case, $\omega_3(\vp) \approx  |\vp|S v_+v_-/\tilde v_F$,  which is equal to the lowest-energy mode of the Mattis model \cite{dWL2}, whereas in the latter case, $\omega_3(\vp) \approx |\vp|^2S c v_-/\Theta$ for $|\vp|<\Theta/c$. Thus the photons can affect the low temperature  thermodynamical properties of the system, no matter how small $|\eR|\neq 0$. For example, we found that the temperature ($T$) dependence of the heat capacity at low $T$ is linear for $\eR=0$: $C_v\propto T $, but there are logarithmic corrections for $\eR\neq 0$: $C_v \propto T\ln(T_0/T)$ with some computable constant $T_0$.

We also studied the magnetic field response to an external current $J^\mu$ with $J^0=0$, i.e., we computed the linear response function $\hat K$ in the relation 
\begin{equation}
\langle \hat B(\omega,\vp)\rangle = \hat K(\omega,\vp)|\vp|\hat J_T(\omega,\vp)
\end{equation} 
with $|\vp|\hat J_T$  the Fourier transform of $\partial_+J_--\partial_-J_+$. We obtain the following exact result
\begin{equation} 
\label{hat K} 
\hat K(\omega,\vp)= \frac{\omega_+^2(\omega_+^2-\Theta^2-\tilde v_F^2|\vp|^2) + v_-^2 S^2|\vp|^2(\Theta^2+v_+^2|\vp^2|)}{(-\omega_+^2+\omega_1(\vp)^2)(-\omega_+^2+\omega_2(\vp)^2)(-\omega_+^2+\omega_3(\vp)^2)}
\end{equation} 
with $\omega_+^2\equiv (\omega + \ii 0^+)^2$ and $\omega_j(\vp)$ the dispersion relations given above.  Expanding $\hat K$ in partial fractions, inserting the long-distance approximations of $\omega_j(\vp)$ given above, and transforming to position space we obtain, for $c\gg v_F$,
\begin{equation} 
\label{Meissner} 
(\partial_t^2 - c^2\nabla^2 +\Theta^2)\langle B \rangle \approx  \partial_+ J_- -\partial_- J_+,\quad B= \partial_+ A_- -\partial_- A_+
\end{equation} 
(see Section~\ref{App_Meissner2} for details). This proves that there is a Meissner effect (in the sense of the London phenomenological theory of superconductivity; see e.g.\ \cite{Tinkham}, Sections~1.2 and 2.1) with a London penetration depth 
\begin{equation} 
\lambda_L = \frac{c}{\Theta} = \frac{c}{e_0}\sqrt{\frac{\pi\ta}{v_F(1-\gamma_1)}} . 
\end{equation} 
It is worth noting that we establish the Meissner effect by a computation that is manifestly gauge invariant (see Appendix~\ref{App_Meissner3} for details). 

The Meissner effect is often regarded as a hallmark of superconductivity (see e.g.\ \cite{Tinkham}). It thus is important to stress that we cannot conclude from our results that our model describes a superconductor: our model is not isotropic and thus quite different from conventional models where Meissner effect is known to correspond to superconductivity. A more detailed investigation of this question would be interesting but is left to future work. 

\section{Conclusions}
\label{sec5} 
The generation of mass in quantum gauge theories has been an important issue both in particle- and condensed matter physics since a long time (see e.g.\ \cite{Schw2,Anderson,H}). The model presented in this paper is, to our knowledge, the first example of an exactly solvable gauge theory in higher dimensions than 1+1 where a gauge field becomes massive by a mechanism similar to the dynamical mass generation in the Schwinger model \cite{Schw3}. This is different from the well-known Higgs mechanism \cite{H} in that no Higgs field is involved; instead, the non-zero photon mass arises due to a commutator anomaly \cite{Schw1}. 

The relations between photon mass generation, Meissner effect, and superconductivity have been a challenging topic in theoretical physics because it is difficult to convincingly reconcile the approximations underlying BCS-theory with gauge invariance (see \cite{Anderson} and references therein). We believe that the exact solution of our model can shed new light on these relations. It would also be interesting to confront our model with model-independent results on superconducting electrodynamics \cite{Sewell}. 

As already mentioned, one motivation for our work is the possible violation of Landau's Fermi liquid theory in models of strongly interacting fermions. This has been an actively researched problem since the discovery of the cuprate high-temperature superconductors in 1986 \cite{BednorzMueller}, and the realization that these materials display many properties not described by Fermi liquid theory \cite{Bonn}. Early on, it was suggested that models of Hubbard-type capture the strongly correlated physics of cuprates \cite{AndersonRVB,Emery,VSA,ZhangRice}. However, it has been proven very difficult to do reliable computations for two dimensional (2D) such models relevant in this context. This situation is very different from the one for 1D correlated fermions which by-now are very well understood. Few would deny that exactly solvable models have played a key role in developing this understanding; we mention the exact solutions of the Luttinger model \cite{MattisLieb} and the 1D Hubbard model \cite{LiebWu}. We thus hope that exactly solvable models will also prove useful to obtain a better understanding of 2D correlated fermion systems. 

It has been known for quite some time that fermions coupled to dynamical photons can have non-Fermi-liquid behaviour \cite{Holstein,Reizer1}. One argument against this mechanism being relevant for real materials is the smallness of the fine-structure constant ($\alpha\approx \frac{1}{137}$), which governs the strength of interactions between matter and transverse photons. Still, various scenarios have been proposed and explored in which effective photon-like gauge fields arise in the low-energy limit of models for strongly correlated fermions \cite{NL,FranzTesanovic:2001,Leeetal}; see \cite{Tsvelik} for review. In these instances, the effective coupling constant need not be small. The computations to explore this mechanism in 2D are usually based on approximations that are difficult to justify and, again,  things are much better understood in 1D due to the existence of exactly solvable prototype models. For example, the (1+1)D quantum gauge theory obtained by minimally coupling the Luttinger model to dynamical photons is exactly solvable \cite{GLR}. This model is a generalization of the Schwinger model \cite{Schw3} and, as the latter, describes photons that are massive. 

Our results suggest the following which could be relevant for the issues discussed in the two previous paragraphs: In derivations of the 2D Hubbard model from more fundamental models of non-relativistic electrons and ions coupled to a dynamical electromagnetic field, only the Coulomb interaction terms are taken into account, while the dynamical transverse photons are ignored. As mentioned, this is usually justified by the smallness of the coupling constant $e_0$ for transverse photons. However, we find that in an effective model of a system of this kind, the bare coupling constant $e_0$ is renormalized, and the renormalized coupling constant $\eR$ can be significant even if the bare one is very small. Moreover, it is possible to study the effects of dynamical photons on 2D Hubbard-like systems away from half filling in an exactly solvable model, and one such effect might be superconductivity. We find it intriguing that the latter conjecture can be proved (or disproved) by exact computations. However, these computations are far from easy and thus left to future work. 

We regard the model in the present paper as a prototype, rather than a candidate for a realistic model of the cuprates: If the gauge field in this model is to be interpreted as a physical electromagnetic field, one should extend it to three dimensions. If the model is to describe correlated fermions in real materials, one should introduce additional spin degrees of freedom. Exactly solvable extensions of our model addressing these remarks exist; we plan to present and study them elsewhere. 

\bigskip

\subsection*{Acknowledgments} 
Useful discussions with F.~H.~L. Essler, L.~B. Ioffe, V.~M. Krasnov, J. Lidmar,  G.~W. Semenoff,  A. Sudb{\o}, M.~Wallin, L.~C.~R. Wijewardhana, and K.  Zarembo are acknowledged. This work was supported by the G\"oran Gustafsson Foundation and the Swedish Research Council (VR) under  contract no. 621-2010-3708.

\appendix

\newcommand{\newsection}{\setcounter{equation}{0}\section}
\renewcommand{\theequation}{\thesection.\arabic{equation}}
\renewcommand{\appendix}{\setcounter{equation}{0}\setcounter{section}{0}\renewcommand{\thesection}{\Alph{section}}}
\newcommand{\appsection}[1]{\setcounter{equation}{0}\renewcommand{\thesection}{\Alph{section}}
\section{#1} \renewcommand{\thesection}{\Alph{section}}}

\newsection{Formal derivation of the model}
\label{App_formal}
We give details on the formal derivation of our model, using the standard canonical formalism; see e.g.\ \cite{Sundermeyer} for further details. (By "formal" we mean here that normal ordering- and regularization issues are ignored.)

We consider the case without direct fermion-fermion interactions (the interacting case follows trivially). The formal Lagrangian density corresponding to the Hamiltonian in \Ref{Mattis Hamiltonian}, for $g_{r,s,r',s'}=0$, is
\begin{equation}
\mathcal{L}_0 = \sum_{r,s=\pm}\Bigl( -\psi^\dag_{r,s}(-\ii\partial_t)\psi^{\phantom\dag}_{r,s}  - v_F\psi^\dag_{r,s}r(-\ii\partial_s)\psi^{\phantom\dag}_{r,s} \Bigr) 
\end{equation}
with $\partial_t\equiv\partial/\partial t$.  We couple this Lagrangian to a dynamical Abelian gauge field by the minimal substitution $-\ii\partial_\mu\to -\ii\partial_\mu + \eB A_\mu$, and by adding the usual Maxwell term $-(c^2/4)F^{\mu\nu}F_{\mu\nu}$, $F_{\mu\nu}=\partial_\mu A_\nu-\partial_\nu A_\nu$ ($\partial_\mu\equiv\partial/\partial x^\mu$; our conventions are as in \cite{Sundermeyer}).  This yields the following Lagrangian density, 
\begin{equation}
\label{cL} 
\mathcal{L} = \sum_{r,s=\pm}\Bigl( -\psi^\dag_{r,s}(-\ii\partial_t+c\eB A_0)\psi^{\phantom\dag}_{r,s}  - v_F\psi^\dag_{r,s}r(-\ii\partial_s+\eB A_s)\psi^{\phantom\dag}_{r,s} \Bigr)+\frac{c^2}{2}\Bigl(F_{0+}^2+F_{0-}^2-B^2 \Bigr)  
\end{equation} 
with $B\equiv F_{+-}$. The canonical momenta $\Pi_X=\partial\mathcal{L}/\partial(\partial_t X)$ ($X=A_\mu$, $\psi\pdag_{r,s}$, $\psi^{\dag}_{r,s}$) are  
\begin{equation}
\Pi_{A_s}= cF_{0s}\equiv E_s,\quad \Pi_{A_0}=0,\quad \Pi_{\psi\pdag_{r,s}}=-\ii\psi^\dag_{r,s},\quad \Pi_{\psi^\dag_{r,s}}=0
\end{equation} 
(the minus sign is due to our conventions for Grassmann derivatives; see \cite{Sundermeyer}, Appendix C).  This yields canonical (anti-) commutator relations for the fermion- and photon field operators as stated in the main text. The Hamiltonian density is defined as $\mathcal{H} = \sum_X (\partial_t X)\Pi_X -\mathcal{L}$, and one finds 
\begin{equation}
\label{calH}
\mathcal{H} = \frac{1}{2} (E_+^2 + E_-^2) + \frac{c^2}{2}B^2  + \sum_{r,s=\pm}v_F\psi^\dag_{r,s}r(-\ii\partial_s+\eB A_s)\psi\pdag_{r,s} + cA_0 G + \sum_{s=\pm}\partial_s(cA_0E_s) 
\end{equation} 
with
\begin{equation}
\label{formalG}
G=-\sum_{s=\pm}\partial_s E_s + \sum_{r,s=\pm} \eB \psi^\dag_{r,s}\psi\pdag_{r,s}.  
\end{equation} 
The primary constraint $\Pi_{A_0}=0$ implies Gauss'  law $G=0$, as usual. With that we obtain the formal Hamiltonian $H=\int \ud^2 x\, \mathcal{H}$ equal to the one in \Ref{Gauged Mattis Hamiltonian} for $g_{r,s,r',s'}=0$, up to normal ordering (we drop the surface term, as usual; note that $E_s=\partial_tA_s-c\partial_sA_0$). 

\newsection{Gauge-invariant normal ordering}
\label{App_normalorder}
We give details on how to derive \Ref{JrsA} and \Ref{gauge-invariant fermion normal-ordering with derivative}.   

Let $\ve_\pm$ be the 2D unit vectors such $\vx=x_+\ve_++x_-\ve_-$. Use
\begin{equation}
\label{we use 1}
\left\langle\psi^\dag_{r,s}(\vx-\epsilon\ve_s/2)  \psi^{\phantom\dag}_{r,s}(\vx+\epsilon\ve_s/2)
 \right\rangle = r\frac1{2\pi\ii \tilde a \epsilon}
\end{equation} 
(see (4.9) in \cite{dWL2}) to define
\begin{equation}
\label{normalJ}
:\!\psi^\dag_{r,s}(\vx)  \psi^{\phantom\dag}_{r,s}(\vx)\!:\,  \equiv \lim_{\epsilon\to 0} \Biggl( \psi^\dag_{r,s}(\vx-\epsilon\ve_s/2)  \psi^{\phantom\dag}_{r,s}(\vx+\epsilon\ve_s/2) -  r\frac1{2\pi\ii \tilde a \epsilon}\Biggr) . 
\end{equation}
This corresponds to Schwinger's point splitting method without gauge field. The singular part does not depend on the gauge field, and therefore the gauge invariant version of \Ref{normalJ} is (note that  both terms in the parenthesis below are gauge invariant) 
\begin{equation}
\label{normalJA}
\nna\psi^\dag_{r,s}(\vx)  \psi^{\phantom\dag}_{r,s}(\vx)\nne\,   \equiv \lim_{\epsilon\to 0} \Biggl( \psi^\dag_{r,s}(\vx-\epsilon\ve_s/2)  \psi^{\phantom\dag}_{r,s}(\vx+\epsilon\ve_s/2)\ee^{\ii\eB\int_{-\epsilon/2}^{\epsilon/2}\ud\xi A_s(\vx+\xi\ve_s) } -  r\frac1{2\pi\ii \tilde a \epsilon}\Biggr) . 
\end{equation} 
To compute this we conclude from \Ref{normalJ} that
\begin{equation*}
 \psi^\dag_{r,s}(\vx-\epsilon\ve_s/2)  \psi^{\phantom\dag}_{r,s}(\vx+\epsilon\ve_s/2) = r\frac1{2\pi\ii \tilde a \epsilon} +J_{r,s}(\vx) + O(\epsilon) 
\end{equation*} 
where $O(\epsilon)$ are terms vanishing in the limit $\epsilon\to 0$. With that, 
\begin{equation*}
\begin{split}
\psi^\dag_{r,s}(\vx-\epsilon\ve_s/2)  \psi^{\phantom\dag}_{r,s}(\vx+\epsilon\ve_s/2)\ee^{\ii\eB\int_{-\epsilon/2}^{\epsilon/2}\ud\xi A_s(\vx+\xi\ve_s) }  = \Bigl( r\frac1{2\pi\ii \tilde a \epsilon} +J_{r,s}(\vx) + O(\epsilon) \Bigr)\\ \times \Bigl(1+\ii\eB\epsilon A_s(\vx) + O(\epsilon) \Bigr) =   r\frac1{2\pi\ii \tilde a \epsilon} +J_{r,s}(\vx)  + r\frac{\eB}{2\pi\tilde{a}}A_s(\vx) + O(\epsilon) . 
\end{split} 
\end{equation*} 
Inserting this in \Ref{normalJA} we obtain \Ref{JrsA}. 

To find the gauge-invariant normal ordering of $\psi_{r,s}^\dag(-\ii\partial_s)\psi_{r,s}$ we start from 
\begin{equation*}
\label{start} 
\psi^\dag_{r,s}(\vx)\psi^{\phantom\dag}_{r,s}(\vx+\epsilon\ve_s) =\, :\! \psi^\dag_{r,s}(\vx)\psi^{\phantom\dag}_{r,s}(\vx+\epsilon\ve_s)\!:\, + r\frac1{2\pi\ii \tilde a \epsilon} .
\end{equation*} 
Differentiating this w.r.t.\ $\epsilon$ and multiplying with $-\ii r$ we obtain
\begin{equation*}
\label{start1} 
\psi^\dag_{r,s}(\vx)r(-\ii\partial_s) \psi^{\phantom\dag}_{r,s}(\vx+\epsilon\ve_s) =\, : \!\psi^\dag_{r,s}(\vx)r(-\ii\partial_s)\psi^{\phantom\dag}_{r,s}(\vx)\!:\, + \frac1{2\pi\tilde a \epsilon^2} +O(\epsilon)
\end{equation*} 
where we use that the limit $\epsilon\to 0$ of the normal ordered fermion bilinear expression is well-defined. 
Thus $\nna \psi^\dag_{r,s}(\vx)r(-\ii\partial_s+\eB A_s(\vx) ) \psi^{\phantom\dag}_{r,s}(\vx)\nne\,$ can be computed as 
\begin{equation}
\label{nn2nd}
\begin{split} 
\lim_{\epsilon\to 0} \Biggl( \ee^{\ii\eB\int_0^{\epsilon}\ud\xi A_s(\vx+\xi\ve_s) } \psi^\dag_{r,s}(\vx)r(-\ii\partial_s+\eB A_s(\vx) ) \psi^{\phantom\dag}_{r,s}(\vx+\epsilon\ve_s) -  \frac1{2\pi\tilde a \epsilon^2}  \Biggr). 
\end{split} 
\end{equation} 
By straightforward computations one obtains (note that the first term in the  parenthesis above can be computed as $ -\ii r(d/ d\epsilon) \psi^\dag_{r,s}(\vx)\psi^{\phantom\dag}_{r,s}(\vx+\epsilon\ve_s)\exp({\ii\eB\int_0^{\epsilon}\ud\xi A_s(\vx+\xi\ve_s) }$)) 
\begin{equation} 
\begin{split} 
\nna \psi^\dag_{r,s}(\vx)r(-\ii\partial_s+\eB A_s(\vx) ) \psi^{\phantom\dag}_{r,s}(\vx)\nne\, = &\, :\! \psi^\dag_{r,s}(\vx)r(-\ii\partial_s+\eB A_s(\vx))\psi^{\phantom\dag}_{r,s}(\vx)\!:\, \\ &+ \frac{\eB^2}{4\pi\tilde{a}}A_s(\vx)^2 + \frac{\eB}{4\pi\tilde{a}} \ii \partial_sA_s(\vx) . 
\end{split} 
\end{equation} 
This implies \Ref{gauge-invariant fermion normal-ordering with derivative}. 

\newsection{Gauge invariance} 
\label{App_gauge} 
The non-trivial gauge transformations at fixed time $t=0$ are 
\begin{equation}
\label{GaugeTrafo}
A_s\to A_s + \partial_s\chi,\quad \psi\pdag_{r,s}\to \ee^{-\ii e_0\chi}\psi\pdag_{r,s},\quad \psi^\dag_{r,s}\to \ee^{\ii e_0\chi}\psi^\dag_{r,s},\quad J_{r,s}\to J_{r,s} -r\frac{e_0}{2\pi\tilde{a}}\partial_s\chi
\end{equation} 
with $\chi\equiv\chi(\vx)$ arbitrary differentiable functions (since any operator $X$ transforms like $X\to X+\ii[G[\chi],X]$ with the Gauss law operators in \Ref{Gauss law generator}).  

With that one can check that  \Ref{JrsA} and \Ref{gauge-invariant fermion normal-ordering with derivative} are gauge invariant as follows: in the former case, this is equivalent to 
\begin{equation}
\label{cJ}
\mathcal{J}_{r,s}\equiv J_{r,s} + r\frac{e_0}{2\pi\tilde{a}}A_s
\end{equation} 
being gauge invariant, and this is a simple consequence of \Ref{GaugeTrafo}. In the latter case, this is seen from 
\begin{equation}
\label{R2}
\begin{split} 
\int\ud^2x\, \nna\psi^\dag_{r,s}(-\ii\partial_s+\eR A_s) \psi^{\phantom\dag}_{r,s}\nne \, = \int\ud^2 x\,  \pi\tilde a \xxa \mathcal{J}_{r,s}^2 \xxe , 
\end{split}
\end{equation} 
which is a simple consequence of \Ref{Kronig identity} and \Ref{cJ}.

\newsection{Exact dispersion relations} 
\label{App_solution} 
We give details on how we obtained the results in \Ref{cp}--\Ref{omega3}. 

The gauge-fixed Hamiltonian in Fourier space mentioned in the main text is 
\begin{equation} 
\label{h3}
\begin{split} 
H_{g.f.} = \frac12 \int \ud^2p\xxa\Biggl( |\hat E_T|^2+\eR^2|\hat \Phi_L|^2  +v_F(1-\gamma_1)|\hat\Pi_L|^2 + v_F(1-\gamma_1)|(\hat\Pi_T-\eR\hat A_T)|^2 \\ + c^2 |\vp|^2|\hat A_T|^2 + v_F(1+\gamma_1)|\vp|^2|\hat\Phi_L|^2 + v_F[\gamma_2-(1+\gamma_1)]p_+p_-( \hat\Phi_+^\dag \hat\Phi_-+h.c.)\Biggr)\xxe
\end{split}
\end{equation}
with $\hat X\equiv \hat X(\vp)$,  $\hat X^\dag\equiv \hat X(-\vp)$, and $|\hat X|^2= \hat X^\dag\hat X$ ($X=A_Y$, $E_Y$, $\Phi_Y$, $\Pi_Y$, and $Y=L,T$). We find it convenient to introduce the following matrix notation
\begin{equation}
\mathbf{P}=\left(\begin{array}{c}\hat P_1\\\hat P_2\\ \hat P_3\end{array}\right)\equiv \left(\begin{array}{c}\hat E_T\\ \hat \Phi_T\\ \hat \Phi_L\end{array}\right),\quad \mathbf{Z}=\left(\begin{array}{c}\hat Z_1\\\hat Z_2\\ \hat Z_3\end{array}\right)\equiv \left(\begin{array}{c}\hat A_T\\ -\hat \Pi_T\\ -\hat \Pi_L\end{array}\right)
\end{equation} 
allowing us to write 
\begin{equation} 
\label{general}
H_{g.f.}=\int \ud^2 p\, \xxe \mathbf{P}^\dag \mathbf{A}\mathbf{P} +  \mathbf{Z}^\dag \mathbf{B}\mathbf{Z} \xxa
\end{equation} 
with the symmetric matrices 
\begin{equation}
\label{AB} 
\mathbf{A}=\left(\begin{array}{ccc} 1 & 0& 0\\ 0& v_2S^2|\vp|^2 & -v_2SC |\vp|^2\\ 0& -v_2SC|\vp|^2 &(v_1-v_2S^2)|\vp|^2+\eR^2 \end{array} \right),\quad \mathbf{B}=\left(\begin{array}{ccc} c^2|\vp|^2+v_3\eR^2 & -v_3\eR & 0 \\  -v_3\eR & v_3 & 0\\ 0 & 0 & v_3 \end{array} \right); 
\end{equation} 
we use the notation 
\begin{equation} 
S\equiv \sin(2\varphi),\quad C\equiv\cos(2\varphi)
\end{equation} 
with $\varphi$ the polar angle in Fourier space (i.e.\ $p_+=|\vp|\cos(\varphi)$, $p_-=|\vp|\sin(\varphi)$), and
\begin{equation} 
\label{vj} 
v_1\equiv v_F(1+\gamma_1),\quad v_2\equiv \frac12 v_F[(1+\gamma_1)-\gamma_2],\quad v_3\equiv v_F(1-\gamma_1) . 
\end{equation} 
The field $\hat Z_j$ and $\hat P_j$ obey the canonical commutator relations of bosons in Fourier space, and the Hamiltonian in \Ref{general} can be diagonalized by a Bogoliubov transformation (see e.g.\  \cite{dWL2}, Appendix~C.1 for details). One finds that the exact dispersion relations of the model are identical with the eigenvalues of the matrix $\mathbf{C}\equiv \mathbf{A}^{1/2}\mathbf{B}\mathbf{A}^{1/2}$. We compute the characteristic polynomial of this matrix as $\det(\lambda-\mathbf{A}\mathbf{B})$ and obtain the result in \Ref{cp}. 

To find the zeros of the polynomial in \Ref{cp} we make the ansatz $[(\lambda-\Theta)^2-A(\lambda-\Theta)-B](\lambda-\lambda_3)$ and compute $A$, $B$ and $\lambda_3$ as power series in $|\vp|^2$ (note that $A=B=\lambda_3=0$ for $|\vp|=0$). In this way we find the boson dispersion relations
\begin{equation}
\label{om1om2}
\begin{split}
&\omega_{1,2}(\vp)^2= \Theta^2 + \frac{1}{2}\Bigl( c^2 + \tilde v_F^2\pm(c^2-\tilde v_F^2 )\sqrt{1 + 4\Gamma S^2}  \Bigr) |\vp|^2 + O(S^2 |\vp|^4)\\
&\omega_3(\vp)^2 =  \frac{c^2v_-^2S^2}{\Theta^2}|\vp|^4 + O(S^2 |\vp|^6)
\end{split} 
\end{equation} 
with $\Gamma\equiv v_-^2[c^2+v_-^2-\tilde v_F^2] /(c^2-\tilde v_F^2)^2$ (note that $v_-^2=v_2v_3$, $\tilde v_F^2 =v_1v_3$,  and $\Theta^2=v_3\eR^2$). This shows that, for $c^2\gg v_F^2$, the gapped modes are approximately rotation invariant, as stated in the main text. The approximation for $\omega_3(\vp)$ above can be improved by noting that the equation determining the zeros of the characteristic polynomial in \Ref{cp} can be written as 
\begin{equation}
\label{om3}
\lambda = |\vp|^4S^2 \frac{v_-^2 ( v_+^2(c^2|\vp|^2-\lambda)+c^2\Theta^2 )}{(\Theta^2+c^2|\vp|^2-\lambda)(\Theta^2 +\tilde v_F^2 |\vp|^2-\lambda)} . 
\end{equation} 
Solving this by iteration starting from $\lambda=0$ yields \Ref{omega3}. 

\newsection{Lagrange formalism}
\label{App_Lagrange}
Many correlation functions of the gauged Mattis model can be computed efficiently using a functional integral formalism. For this one needs  the Lagrange formalism of the bosonized model.

The formal\footnote{As before, we mean by this that regularization- and normal ordering issues are ignored.} Lagrangian density corresponding to our bosonized model is 
\begin{equation}
\label{cLGM}  
\begin{split} 
\mathcal{L} = \frac12\sum_{s=\pm}(\partial_tA_s-c\partial_sA_0) ^2 + \sum_{s=\pm} \Bigl( \frac1{2v_3} ( \partial_t\Phi_s)^2  + \eR A_s\partial_t\Phi_s - c\eR  A_0 \partial_s\Phi_s\Bigr) \\ -\frac{c^2}2(\partial_+A_--\partial_-A_+)^2 -\frac{v_F}2(1+\gamma_1)\sum_{s=\pm}(\partial_s\Phi_s)^2 -v_F\gamma_2(\partial_+\Phi_+)(\partial_-\Phi_-)
\end{split} 
\end{equation} 
with $v_3$ in \Ref{vj} (this can be easily checked by going through the canonical formalism, similarly as in Appendix~\ref{App_formal}:  compute the canonical momenta $\Pi_X=\partial\mathcal{L}/\partial(\partial_t X)$, $X=A_s$, $\Phi_s$, $A_0$, and the Hamiltonian $H=\int \ud^2 x(\sum_X \Pi_X\partial_t X-\mathcal{L})$, and verify that this leads to the Hamiltonian and Gauss' law constraint in \Ref{HGM} and \Ref{bG}).  

It is instructive to check gauge invariance: Gauge transformations are given by
\begin{equation}
A_s\to A_s+\partial_s\chi,\quad A_0\to A_0+\frac1c\partial_t\chi ,\quad \Phi_s\to\Phi_s
\end{equation} 
with arbitrary differentiable functions $\chi(t,\vx)$, and they transform the Lagrangian in \Ref{cLGM} as follows,
\begin{equation*} 
\mathcal{L}\to \mathcal{L} + \eR\sum_{s=\pm} \Bigl( (\partial_s\chi)(\partial_t\Phi_s) -(\partial_t\chi)(\partial_s\Phi_s) \Bigr) = \mathcal{L} + \eR\sum_{s=\pm} \Bigl( \partial_s( \chi \partial_t\Phi_s) -\partial_t(\chi \partial_s\Phi_s) \Bigr),
\end{equation*} 
i.e.\ there is a change by a surface term (which, for our purposes, can be ignored). 

For completeness we give the components of the gauge current $j^\mu \equiv \partial \mathcal{L}/\partial A_\mu$, 
\begin{equation} 
j^0 = -c\eR\sum_s\partial_s\Phi_s,\quad j^\pm = \eR\partial_t\Phi_\pm
\end{equation} 
which obey the continuity equation $\partial_\mu j^\mu=0$, as they should.  Moreover, the Euler-Lagrange equations for the Lagrangian in \Ref{cLGM} are the 2+1 dimensional Maxwell equations 
\begin{equation}
\begin{split} 
&\partial_t E_\pm \pm c^2 \partial_\mp B = j^\pm\\
&\partial_+E_+ + \partial_-E_- = -j^0/c\\
&\partial_t B -\partial_+E_-+\partial_-E_+ =0
\end{split} 
\end{equation} 
($E_\pm = \partial_t A_\pm-c\partial_sA_0$ and $B=\partial_+A_--\partial_-A_+$, as before), and the Klein-Gordon-type equations 
\begin{equation} 
\Bigl( \frac1{(1-\gamma_1^2) v_F^2}\partial_t^2 -\partial_\pm^2\Bigr)\Phi_\pm -\frac{\gamma_2}{1+\gamma_1}\partial_+\partial_-\Phi_\mp +\frac{\eR}{v_F(1+\gamma_1)}E_\pm=0. 
\end{equation}

\newsection{Meissner effect computation}
\label{App_Meissner} 
We give details on how we obtained the result in \Ref{hat K}. 

\subsection{Functional integral formalism} 
\label{App_Meissner1} 
We start with the Minkowski action defined by the Lagrangian in \Ref{cLGM}, perform a Wick rotation $\ii t\to \tau$, $A_0\to \ii A_0$, change the sign of the action, and add a gauge fixing term to obtain  the Euclidean action $S_E=\int_0^\beta \ud\tau\int \ud^2 x\mathcal{L}_E$ with 
\begin{equation}
\label{cLGM1}  
\begin{split} 
\mathcal{L}_E = \frac12\sum_{s=\pm}(\partial_\tau A_s-c\partial_s A_0) ^2 + \sum_{s=\pm} \Bigl( \frac1{2v_3} ( \partial_\tau\Phi_s)^2  - \eR A_s\ii \partial_\tau\Phi_s + \ii c\eR  A_0 \partial_s\Phi_s\Bigr) \\ +\frac{c^2}2(\partial_+A_--\partial_-A_+)^2 +\frac{v_F}2(1+\gamma_1)\sum_{s=\pm}(\partial_s\Phi_s)^2 +v_F\gamma_2(\partial_+\Phi_+)(\partial_-\Phi_-)\\
+\frac{1}{2\xi}(\partial_\tau A_0 + c\partial_+A_++ c\partial_-A_-)^2
\end{split} 
\end{equation} 
and inverse temperature $\beta$ (we perform the computation in a manifestly gauge invariant way using the $R_\xi$-gauge; see e.g.\  \cite{W}). We do a Fourier transform with respect to space and time, and we write the resulting Lagrangian in matrix form as follows, 
\begin{equation}
\mathcal{\hat L}_E = \frac12 \left(\begin{array}{c} \hat A^\dag \, , \, \hat\Phi^\dag \end{array}\right) \left(\begin{array}{cc} D & E^\dag \\ E & F\end{array} \right)\left(\begin{array}{c} \hat A\\ \hat\Phi\end{array}\right) ,
\end{equation} 
with 
\begin{equation}
\hat A=\left(\begin{array}{c} \hat A_0\\\hat A_+\\ \hat A_-\end{array}\right),\quad \hat\Phi=\left(\begin{array}{c} \hat\Phi_+\\ \hat\Phi_-\end{array}\right) 
\end{equation} 
and
\begin{equation}
\label{DEF} 
\begin{split} 
D = \left(\begin{array}{ccc} 
c^2|\vp|^2 & -c\omega p_+ & -c\omega p_-\\
-c\omega p_+ &\omega^2 + c^2p_-^2 & -c^2 p_+p_- \\
-c \omega p_- & -c^2 p_+p_- &\omega^2 + c^2p_+^2
\end{array} \right) +\xi \left(\begin{array}{ccc} \omega^2 & \omega cp_+&\omega cp_-\\
\omega c p_+ & c^2 p_+^2 & c^2 p_+p_- \\
\omega c p_-& c^2 p_+ p_-& c^2 p_-^2
 \end{array} \right) \\
E=\left(\begin{array}{ccc} -\eR cp_+& \eR\omega& 0\\ -\eR cp_-&0&\eR\omega \end{array}\right), \quad F=\left(\begin{array}{cc} \omega^2/v_3+v_+p_+^2&v_F\gamma_2p_+p_-\\ v_F\gamma_2 p_+p_-& \omega^2/v_3+v_+p_-^2 \end{array} \right)
\end{split} 
\end{equation} 
(we find it convenient to collect the gauge fields $\hat A$ and the matter fields $\hat\Phi$ in separate groups). In the following we use the following notation: $p$ is short for $(\omega/c,p^+,p_-)$; $\int \ud^3p$ is short for $(1/\beta)\sum_\omega\int \ud^2p$ with $\omega$ boson Matsubara frequencies as usual; functional integrals are symbolically written as $\int D[A]$ (integration over the gauge fields) and $\int D[\Phi]$ (integration over the matter fields). 

We compute response functions using the generating functional
\begin{equation} 
Z[J] = \int D[A]\int D[\Phi]\exp\Bigl(-\int d^3p\,\bigl[ \mathcal{\hat L} -\sum_{\mu=0,+,-}\hat J^\dag_\mu(p)\hat A^\mu (p)  \bigr] \Bigr)
\end{equation} 
where $\hat J_\mu^\dag(p)=\hat J_\mu(-p)$ are external currents. It is convenient to first perform the functional integral over $\Phi$, and then the functional integral over $A$. This yields
\begin{equation} 
Z[J] = \exp \left( \hat J^\dag P \hat J/2\right),\quad P= (D-E^\dag F^{-1}E)^{-1}
\end{equation} 
where $F^{-1}$ is the matrix inverse of $F$ etc.\ (our normalization is such that $Z[0]=1$). From this we can deduce all response functions of our model as follows, 
\begin{equation}
\label{AJ} 
\langle \hat A^\mu(p)\rangle = \frac{\partial \ln Z[J]}{\partial \hat J^\mu(-p)} = \sum_{\nu=0,+,-} P_{\mu\nu}(p)\hat J\nu(p). 
\end{equation} 

\subsection{Magnetic field induced by transverse current} 
\label{App_Meissner2}
We use \Ref{AJ} and $\hat B(p)= \ii p_+\hat A_-(p) -\ii p_-\hat A_+(p)$ to compute $\langle \hat B(p)\rangle$. Choosing a transverse external current $\hat J_\mu(p)$, i.e.\ $\hat J_0(p)=0$ and $\hat J_\pm(p) = \pm \ii p_\mp \hat J_T(p)/|\vp|$ for some function $\hat J_T(p)$, we obtain  
\begin{equation} 
\langle \hat B(p)\rangle = K_E(p) \hat J_T(p)  
\end{equation} 
with
\begin{equation}
\label{KE} 
K_E(\omega,\vp) =  \frac{\omega^2(\omega^2+\tilde v_F^2|\vp|^2+\Theta^2)+v_-^2 |\vp|^2S^2 (v_+^2 |\vp|^2+\Theta^2)}{(\omega^2+\omega_1(\vp)^2) (\omega^2+\omega_2(\vp)^2) (\omega^2+\omega_3(\vp)^2)} |\vp|.  
\end{equation} 
The result in \Ref{hat K} is obtained from this by analytical continuation, i.e., 
\begin{equation}
\hat K(\omega,\vp) = \hat K_E(\omega,\vp)|_{\omega\to -\ii\omega +0^+}.
\end{equation} 
Expanding $\hat K$ in partial fractions and inserting the long-distance approximations of $\omega_j(\vp)$ given in the main text, we obtain 
\begin{equation} 
\langle \hat B(\omega,\vp)\rangle  =  \frac{1}{-\omega_+^2+c^2|\vp|^2+\Theta^2} |\vp|\hat J_T (\omega,\vp) + \ldots 
\end{equation} 
with dots indicating terms suppressed by a factor $(v_F/c)^2S^2$. Transforming this to position space we obtain \Ref{Meissner}. 

Equation \Ref{Meissner} suggests that, in the long wavelength limit $(\omega,\vp)\to(0,\mathbf{0})$ and for $v_F\ll c$, $\langle A_\pm\rangle \approx \Theta^{-2} J_\pm$, which are the equations underlying the London phenomenological theory of superconductivity; see e.g.\ \cite{Tinkham}, Sections~1.2 and 2.1. 

\subsection{Manifest gauge invariance}
\label{App_Meissner3} 
In the previous section we computed one response function for our model and obtained a gauge invariant result (i.e.\ independent of the gauge fixing parameter $\xi$). We now prove gauge invariance of the formalism we use in general. 

Physical external currents have to obey the continuity equation $\partial_\mu J^\mu=0$, i.e., $\ii\omega \hat J_0 + \ii p_+\hat J_++\ii p_-\hat J_-=0$, and currents of particular interest are transverse as defined in the previous section. It therefore is useful to introduce orthonormal basis vectors $\he_j(p)$, $j=1,2,3$, such that physical currents are of the form $\hat J_\mu = \hat J_2(\he_2)_\mu + \hat J_3(\he_3)_\mu$ (i.e.\ the continuity equation corresponds to $\hat J_1=0$) with $\hat J_2=\hat J_T$ as defined in the previous section, i.e., 
\begin{equation}
\he_1 = \frac1{\Omega}\left(\begin{array}{c} \ii\omega\\\ii cp_+\\\ii cp_-\end{array}\right),\quad \he_2=\frac1{|\vp|}\left(\begin{array}{c} 0 \\\ii p_-\\-\ii p_+\end{array}\right),\quad \he_3=\frac1{|\vp|\Omega}\left(\begin{array}{c} -c|\vp|^2 \\ \omega p_+\\ \omega p_-\end{array}\right)
\end{equation} 
with $\Omega\equiv\sqrt{\omega^2+c^2|\vp|^2}$ (we choose phase factors such that $(\he_j^\dag(p))_\mu = (\he_j\pdag(-p))_\mu$). We observe that all gauge invariant combinations of $\hat A_\mu$ are given by 
\begin{equation} 
\label{BEA}
\begin{split} 
\hat B= |\vp|\he_2^\dag\cdot\hat A,\quad \hat E_L =-\ii\Omega\he_3^\dag\cdot\hat A, \quad  \hat E_T = \omega\he_2^\dag\cdot\hat A
\end{split} 
\end{equation} 
(we used that $ \hat E_s=-\ii\omega\hat A_s-\ii cp_s\hat A_0$; $\he_j^\dag\cdot\hat A$ is short for $(\he_j^\dag)_\mu \hat A^\mu$). To prove gauge invariance we therefore only have to compute  the matrix $P$ in this basis.  We find 
\begin{equation} 
\label{Pjk} 
(\he_j^\dag \cdot P\he_k)_{j,k=1}^3 = \left(\begin{array}{ccc} P_{11} & 0 & 0\\ 
0 & P_{22} & P_{23} \\ 0 & P_{32} & P_{33} \end{array}\right) 
\end{equation} 
with $P_{11}=1/(\xi \Omega)$ and $P_{jk}$ independent of $\xi$ for $j,k=2,3$ (e.g.\ $P_{22}(\omega,\vp)|\vp|=K_E(\omega,\vp)$ in \Ref{KE} etc.). The block diagonal form of the matrix in \Ref{Pjk}  proves that physical currents can only induce gauge invariant responses, as expected.

\end{document}